\documentclass[fleqn,10pt]{wlscirep}
\title{On the (un)effectiveness of Proton Boron Capture in Proton Therapy}

\author[1]{Annamaria Mazzone}
\author[2]{Paolo Finocchiaro}
\author[3]{Sergio Lo Meo}
\author[4]{Nicola Colonna}
\affil[1]{Consiglio Nazionale delle Ricerche - Istituto di Cristallografia (CNR - IC), Bari, Italy}
\affil[2]{Istituto Nazionale Fisica Nucleare, Laboratori Nazionali del Sud (INFN - LNS), Catania, Italy}
\affil[3]{ENEA, Centro Ricerche "E. Clementel", Bologna, Italy}
\affil[4]{Istituto Nazionale Fisica Nucleare (INFN), Sez. di Bari, Bari, Italy}
\affil[*]{nicola.colonna@ba.infn.it}

\begin{abstract}
We present calculations and simulations on the role of the p+$^{11}$B$\to$3$\alpha$ reaction in proton therapy. This reaction has been recently suggested to be responsible for a decrease in the survival probability of tumor cells, when they are irradiated with low-energy protons. However, at the concentration levels typical of the proposed boron carrier (sodium borocaptate, Na$_{2}$B$_{12}$H$_{11}$SH, in short BSH), i.e. less than 100 ppm, both calculations and Monte Carlo simulations suggest that the dose related to this reaction is orders of magnitude lower than the dose delivered by the primary proton beam inside the tissues. These calculations cast some doubts on the claim of an important role played by Proton Boron Capture in enhancing the therapeutic effectiveness of proton therapy, and suggest that other mechanisms should be investigated in order to explain the observed decrease in the survival probability. 

\end{abstract}
\begin{document}

\flushbottom
\maketitle
% * <john.hammersley@gmail.com> 2015-02-09T12:07:31.197Z:
%
%  Click the title above to edit the author information and abstract
%
\thispagestyle{empty}

\section*{Introduction}

Proton therapy has been widely and successfully used since a few decades for cancer therapy. For deep-seated tumors, the use of high-energy protons allows delivering a higher dose to the tumor, relative to healthy tissues, thanks to the increase of the proton energy loss as they slow down in the matter, in particular at the end of their range, where the so-called Bragg peak is formed. A few years ago, Yoon \textit{et al.} \cite{yoon:2014} suggested that an increase of up to a factor of two of the dose at the proton Bragg peak could be achieved if boron is accumulated in the tumor tissues. The authors based their claim on Monte Carlo (MC) simulations performed with the Monte Carlo N-Particle (MCNP) transport code \cite{mcnp}, assuming an unspecified concentration of $^{11}$B (presumably 100\% over a small boron uptake region, or BUR). The mechanism responsible for a higher dose was suggested to be related to proton-boron fusion reactions, leading to the production of low-energy, and hence high Linear Energy Transfer (LET), $\alpha$-particles. Simulations have also been employed to compare the effect of proton boron fusion therapy (PBFT) with the one of Boron Neutron Capture Therapy (BNCT) \cite{yoon:2017}, concluding on the usefulness of the proposed modality for application in radiotherapy. In this case, results referred in principle to a boron concentration of 1.04 mg/g, although a higher (undeclared) concentration was used in the simulations in order to observe the effect. More recently, Cirrone \textit{et al.} \cite{cirrone:2017} have reported an \textit{in vitro} study showing that the effectiveness of proton therapy is enhanced if natural boron is added in adequate concentration to the tumor cells. In analogy with Boron Neutron Capture Therapy \cite{bnct}, the proposed mechanism, named in Ref. \cite{cirrone:2017} "Proton Boron Capture Therapy" (PBCT), is a two-component treatment modality in which a therapeutic dose is produced by proton-induced reactions on boron, selectively delivered to the tumor by means of a suitable carrier. The effect of PBCT is demonstrated by Cirrone \textit{et al.} \cite{cirrone:2017} by irradiating samples of prostatic tumor cells with and without boron. It is shown that the presence of boron in a concentration of several tens ppm leads to an important decrease in the cell survival probability. Furthermore, by placing the specimens on the surface and at a depth corresponding to the Bragg peak inside a phantom, and irradiating them with 62.5 MeV protons, the authors demonstrate that Proton Boron Capture has an effect only in depth, where low-energy protons are present. Such a feature would naturally lead to an increased selectivity of the PBCT, even without a large gain in boron concentration in tumor tissues relative to healthy ones, a clear advantage of this modality over other two-components treatments.

The authors argue that the reaction responsible for the large dose delivery is the p+$^{11}$B$\to$3$\alpha$. The high Q-value of this reaction, of 8.59 MeV, results in an average energy per alpha particle of 
$\sim$2.9 MeV. As in BNCT, such low-energy $\alpha$-particles are characterized by a range in tissue of a few $\mu$m, comparable to the cell dimension. Furthermore, the high LET that characterizes them can potentially lead to high damage to the cellular DNA. An analysis of the chromosome damage seems to corroborate the hypothesis of the dominant role of the p+$^{11}$B$\to$3$\alpha$ reaction. Another important feature of this reaction is that its cross section shows a resonance-like structure at low proton energy, i.e. 700 keV. As a consequence, the maximum effect of the proton Boron Capture Therapy would be achieved right at the end of the proton range, enhancing the efficacy of the Bragg peak while sparing healthy tissues.

While the experimental evidences of an effect of BSH on the cell survival probability is clear, the statement that such effect could be related to the p+$^{11}$B reactions is mostly speculative. In particular, the article does not report calculations of the reaction rate nor simulations of the dose related to this reaction, to support with solid arguments the claim of an important effect of Proton Boron Capture on Proton Therapy. On the contrary, the previous articles on the subject, by Yoon \textit{et al.} \cite{yoon:2014} and Jung \textit{et al.} \cite{yoon:2017} are based on simulations, but in both cases the employed boron concentration is not clearly specified and only qualitative conclusions are given. 

In order to better understand the potential role of the p+$^{11}$B reaction in boron-enhanced proton therapy, we have performed calculations of the reaction rate and MC simulations of the corresponding dose. The article is organized as follows: the results of the reaction rate calculations and of the GEANT4 \cite{geant4} simulations are shown in the \textit{Results} section, while the implication of the predicted dose is presented in the section \textit{Discussion}. Details on the calculations and on the simulations are given in the section \textit{Methods}.

\section*{Results}

The recent observation of a lower survival probability and higher chromosome damage in a sample of tumor cells in which a boron carrier, BSH, is added prior to proton irradiation, has led Cirrone \textit{et al.} \cite{cirrone:2017} to conclude that a significant dose is delivered to the cells resulting from the p+$^{11}$B$\to$3$\alpha$ reaction. With the aim of verifying the effective role of the reaction in the therapy, and hence of the proposed Proton Boron Capture Therapy, we have performed analytical calculations as well as MC simulations. In the following, we present the results of the calculations, performed on the basis of  some simple considerations, while in a subsequent subsection we report on simulations performed with the simulation toolkit GEANT4, specifically optimized for this case.

\subsection*{Estimate of the p+$^{11}$B$\to$3$\alpha$ dose contribution}

According to Ref. \cite{cirrone:2017}, the maximum effect of PBCT is obtained at the end of the proton range, due to a resonance-like structure peaking around 700 keV. At its maximum, a cross section of $\sim$1 barn has been reported by Segel \textit{et al.} \cite{exfor,segel}. This value allows one to calculate the yield of the p+$^{11}$B $\to$ 3$\alpha$ reactions, Y$_{3\alpha}$, i.e. the probability that a proton undergoes such a reaction, according to the following equation:

\begin{equation}
Y_{3\alpha}= \sigma_{R}\times n_{B}
\label{eq:one}
\end{equation}

Here, $\sigma_{R}$ is the reaction cross section in barns, and n$_{B}$ is the areal density of $^{11}$B, expressed in atoms/barn. Note that this formula is valid in the case of a thin target approximation ($\sigma_{R}\times n_{B}\ll$ 1). In other words, we have neglected here self-absorption corrections that typically would lead to a smaller yield, an assumption that is well justified in this case, as will be shown later on. 

\begin{figure}[h!]
  \includegraphics[width=10.cm]{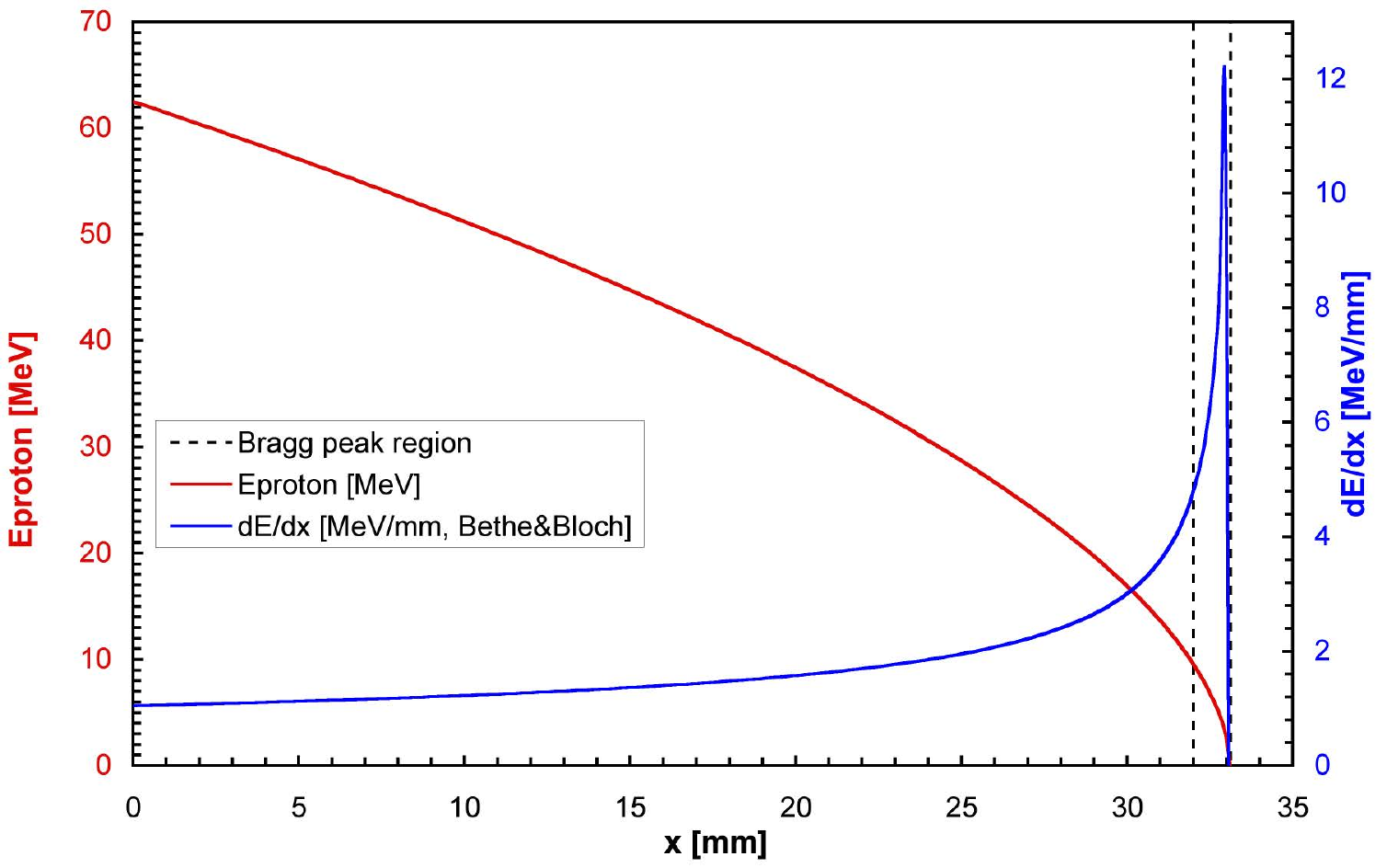}
\caption{The proton dose profile and residual energy as a function of depth inside a phantom, for an incident proton of 62.5 MeV energy. \label{fig:bragg}}
\end{figure}

The value of the areal density to be used in the formula can be determined by considering the thickness of a slab of tissue inside which protons interact. One possibility is to consider a slab of tissue comprising the Bragg peak and assume that the cross section within that region stays constant at its maximum value of $\sim$1. barn. This is clearly a coarse overestimate, if one considers that the cross section is higher than 200 mb only in the range  300 keV$\le$E$_{p}\le$1 MeV (being E$_{p}$ the proton kinetic energy). Nevertheless, for the purpose of this paper, the assumption of a flat reaction cross section in the full range of the Bragg peak is not crucial, if one considers that this is the most favorable scenario for the p+$^{11}$B reaction, and any other assumption would lead to a lower reaction rate. From Figure \ref{fig:bragg}, it can be seen that protons release the last 10 MeV of their kinetic energy in approximately 1 mm of water. Therefore, it seems natural to assume this thickness for the calculation, keeping in mind the approximation mentioned above. Another advantage of this choice  is that one can easily compare a ratio between the total energy released by a proton in the last millimeter of its range, with the total energy released by the p+$^{11}$B reaction, both of which being around 10 MeV. 

Details of the calculations of the $^{11}$B areal density are provided in the "Method" section. For 80 ppm of $^{11}$B, the areal density corresponding to a 1 mm slab of water is 5.5$\times$10$^{-7}$ atoms/barn. Accordingly, and assuming a flat cross section of 1 barn, the 3$\alpha$ reaction yield (i.e. the number of reactions per proton) turns out to be less than 10$^{-6}$, in the best case scenario. Considering that both a proton and the three $\alpha$ particles emitted in the p+$^{11}$B reaction release $\sim$10 MeV in the considered slab, this value represents approximately also the ratio of the doses delivered to the tissue. Clearly, this calculation does not consider the higher biological effectiveness of $\alpha$-particles, relative to protons. Nevertheless, the calculated yield is much too low to affect, in any considerable way, the effectiveness of proton therapy, and casts serious doubts on the conclusion that the decrease of cell survival probability observed by Cirrone \textit{et al.} \cite{cirrone:2017} is related to the p+$^{11}$B$\to$3$\alpha$ reaction.

It can be argued that the choice of the slab thickness affects the proton/3$\alpha$ dose ratio. A much thinner slab would, in fact, result in a lower proton energy release, compared with the fixed energy release in the p+$^{11}$B$\to$3$\alpha$ reaction, deposited over a range of the order of a few microns (comparable with the cell dimension). It should be considered, however, that the choice of a thinner slab leads also to a smaller areal density of boron (i.e. the factor n$_{B}$ in Eq. \ref{eq:one}). To verify whether the two effects (decrease of proton dose and decrease of reaction yield) compensate each other, we have made the same calculations described above in smaller steps, obtained by dividing the region of the Bragg peak in slabs of $\sim$100 $\mu$m thickness each, and assuming as in the previous case, a constant reaction cross section of 1 barn regardless of the proton energy. The results is that in all slabs the energy released by a proton is six orders of magnitude higher than the energy released by the 3$\alpha$ reaction (obtained by multiplying the reaction yield for the Q-value), and this even in the assumption of a cross section fixed at its maximum value, which represents clearly a gross overestimate.
It seems highly unlikely that, in such a condition, a significant enhancement of the therapeutic effect might derive from the so-called Proton Boron Capture Therapy.

It may be useful at this point to spend a few words on another consideration that poses some questions on the interpretation of the important role of the p+$^{11}$B reaction: the presence of proton-induced reactions on $^{16}$O. Several studies have indicated that these reactions contribute to the total dose, as they produce secondary particles, in particular neutrons and $\gamma$-rays (see for example Ref. \cite{neutrons,neutrons2}). Emission of multiple $\alpha$-particles in proton-induced reaction on $^{16}$O have also been reported, with a cross section of the order of 100 mb \cite{exfor, oxy}, i.e. within a factor of ten relative to the maximum value of the p+$^{11}$B$\to$3$\alpha$ reaction. Although the p+$^{16}$O$\to$$\alpha$ reaction occurs above a threshold of 5.5 MeV proton energy, considering that tissues are mostly made of water, one can expect the contribution to the total dose of proton-induced reactions on oxygen to be much higher than the one related to the p+$^{11}$B reaction. Most importantly, the dose related to the $\alpha$-particles from p+$^{16}$O reactions would obviously be distributed almost uniformly over the whole proton range. 

\begin{figure}[h!]
  \includegraphics[width=10.cm]{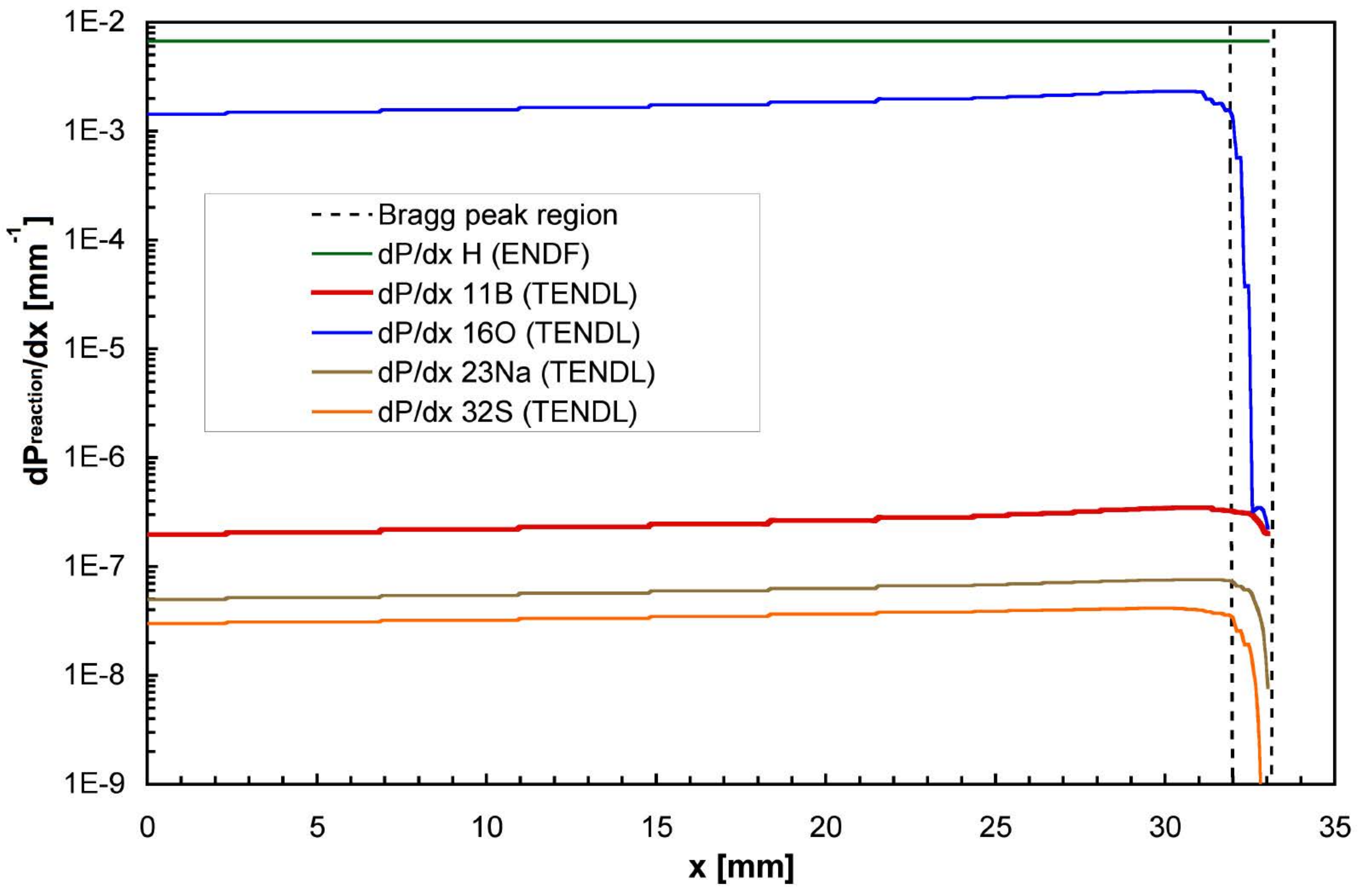}
\caption{The probability (per unit length) of proton interaction as a function of depth inside phantom, for an incident proton beam of 62.5 MeV energy. The different curves represent reactions on the different elements, entering in the composition of the phantom (i.e. water) or of the BSH boron carrier (in 80 ppm).  \label{fig:dpdx}}
\end{figure}

To illustrate this effect, Figure \ref{fig:dpdx} shows the probability per unit length of proton interaction on the tissue constituents, including BSH at the 0.17 mg/ml concentration, corresponding to 80 ppm, as declared in Ref. \cite{cirrone:2017}. For simplicity, we have considered the total reaction cross section, i.e. including elastic and all inelastic reaction channels. It should be noted that, apart from Hydrogen, for which the elastic channel almost entirely dominates the total cross section, for all other elements the elastic cross section contributes for only 10\% to the total cross section. For the inelastic reactions, the cross sections for the various elements are similar, i.e. do not vary by more than one order of magnitude, so does presumably also the energy released by secondary particles. The cross sections used for the calculations shown in Figure \ref{fig:dpdx} are extracted from the library TENDL-2012 \cite{tendl1,tendl2}, which is based on the nuclear interaction model TALYS \cite{talys}. At presents, this is the only evaluated data library that contains cross sections for the p+$^{11}$B interaction, and as shown later on in this paper, it is one of the few options for simulating proton-induced reactions in proton therapy. Although this model is generally considered inaccurate for predicting reaction cross sections on light nuclei, it is well adequate for the purpose of the present paper.

As evident in the figure, the probability of proton interaction in tissue in the presence of BSH is by far dominated by proton-induced reactions on oxygen, with the contribution of boron four orders of magnitude smaller. Although the calculations in the figure are just a crude approximation, that do not take into account the exact value of the ($p, \alpha$) cross section nor the energy loss in the various reactions, the difference between p+$^{16}$O and p+$^{11}$B reaction is so large that it can hardly be compensated by reasonable differences in the cross section, or by the higher energy released in the proton boron capture. The only advantage of the p+$^{11}$B reaction is related to its positive Q-value (i.e. the absence of a reaction threshold), that makes this reaction dominate over the oxygen one in the Bragg peak, as can be clearly appreciated in the figure. In conclusion, it seems evident from the comparison that the small boron concentration and the relatively small cross section of the p+$^{11}$B result in a negligible contribution of this reaction to the total dose, much smaller even than the naturally occurring proton induced reactions on oxygen. 

\begin{figure}[h!]
  \includegraphics[width=17.cm]{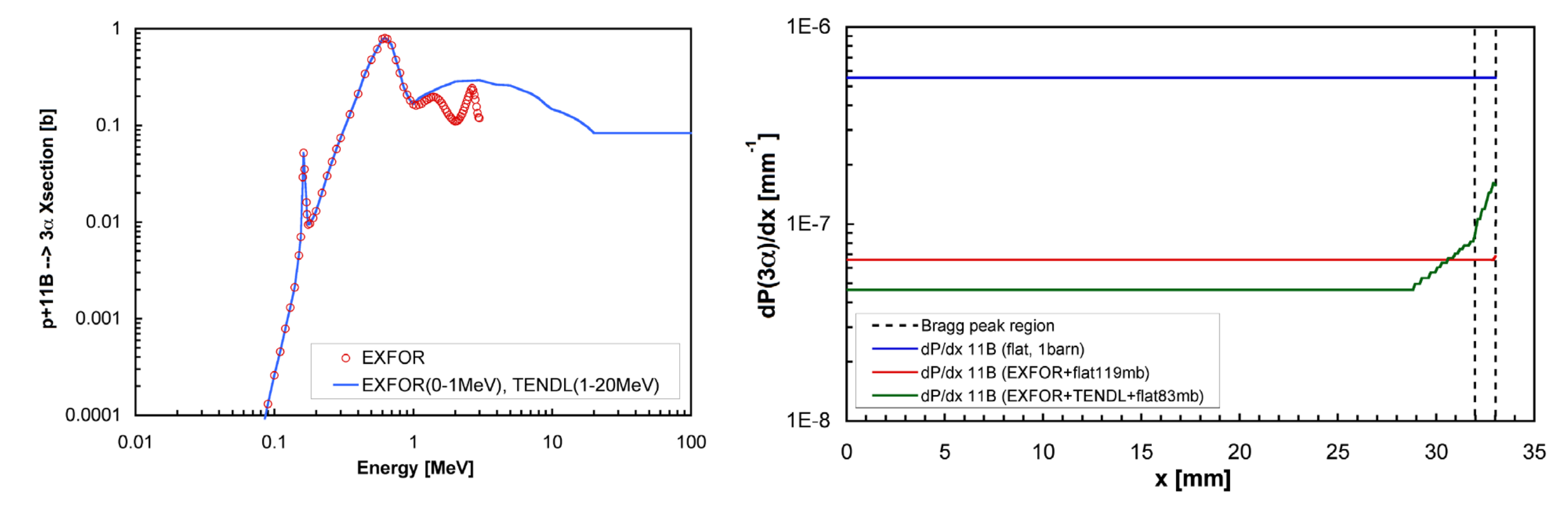}
\caption{Left panel: The cross section of the p+$^{11}$B$\to$3$\alpha$ reaction, as extracted from EXFOR (symbols), and the one contained in the TENDL-2012 evaluated data library, above 1 MeV. The blue curve represents a combination of the experimental and evaluated cross section, used in this work for reaction yield calculations. Right panel: the probability (per unit length) of the p+$^{11}$B$\to$3$\alpha$ reaction as a function of depth inside phantom, for an incident proton beam of 62.5 MeV energy and 80 ppm of boron concentration in tissue. The different curves show the calculations performed with different energy-dependent cross sections: a constant value at all energy of 1 barn (blue curve);  experimental cross section up to 3 MeV, followed by a constant value of 119 mb at higher energy (red curve); a combination of experimental data (for proton energy up to 3 MeV), TENDL-2012 (from 3 to 20 MeV) and a constant value of 83 mb above 20 MeV.  \label{fig:react11B}}
\end{figure} 

While the calculations shown in Figure \ref{fig:dpdx} are made with total cross sections for the various elements, a more refined analysis of the effect of the p+$^{11}$B$\to$3$\alpha$ reaction can be performed by using the experimental cross section from EXFOR \cite{exfor} and/or the evaluated cross section in TENDL-2012 \cite{tendl1,tendl2}, and/or a constant value. The cross sections used in the calculations are shown in Figure \ref{fig:react11B}, left panel. The choice of the cross sections from TENDL-2012 does not dramatically affect the results of the calculations. In the right panel of Figure \ref{fig:react11B} the reaction probability (i.e. the yield calculated with Eq. \ref{eq:one}) is shown for different assumptions on the energy dependence of the p+$^{11}$B$\to$3$\alpha$ cross section, obtained from different combinations of experimental data (up to 3 MeV), evaluated cross sections from TENDL-2012 and constant values. In all cases, the yield is below 10$^{-6}$, a value too low to significantly affect the dose profile in proton therapy.

\begin{figure}[h!]
 \includegraphics[width=17.cm]{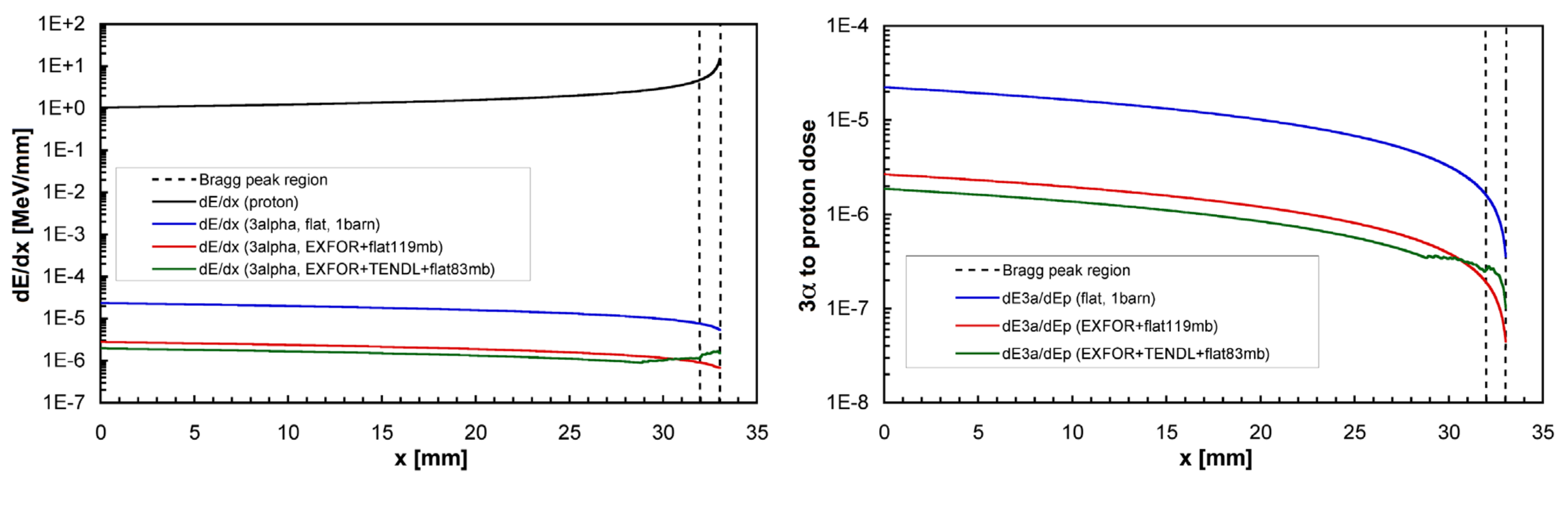}
\caption{Left panel: energy loss of protons of 62 MeV incident energy in water, compared with the energy loss by the $\alpha$ particles produced in the p+$^{11}$B reaction, calculated on the basis of various combinations of the reaction cross section. Right panel: ratio of the proton boron capture and primary proton dose, as a function of the depth inside phantom. \label{fig:relativedose}}
\end{figure}

For a more quantitative assessment of the effect of PBCT, the reaction yield has been used to calculate the dose delivered by the 3 $\alpha$-particles from the p+$^{11}$B reaction, by simply considering the Q-value of the reaction plus the average fraction of the proton kinetic energy transferred to the alpha particles (which could be important for higher energy protons). This dose can be directly compared with the one produced by the primary proton itself, if we neglect the (small) enhancement related to the different relative biological effectiveness for different particles. Figure \ref{fig:relativedose}, left panel, compares the energy loss by the primary protons and the one of $\alpha$-particles from the p+$^{11}$B reaction, in the various assumptions of the cross sections previously mentioned. The ratio between the dose from the proton boron capture and the one due to the primary proton, as a function of the depth inside the tissue, is shown in the right panel of Figure \ref{fig:relativedose}. It can be noted that, apart from being very low in absolute, the depth profile of this ratio is opposite to what would be desirable for enhancing the therapeutic effectiveness of proton therapy, and it is in clear contradiction with the experimental results of Cirrone \textit{et al.} \cite{cirrone:2017}. 

\subsection*{GEANT4 simulations}
 
Monte Carlo simulations are a useful tool to estimate the dose released in hadrontherapy. Typically, simulations of the dose in radiotherapy in general, and proton therapy in particular, are performed with three main tools: MCNP \cite{mcnp}, FLUKA \cite{fluka} and GEANT4 \cite{geant4}. Several studies have compared the performance of the different codes, showing that the general features of the dose profile are well reproduced in all cases, with some small differences related to the treatment of various processes, in particular of the hadronic processes \cite{mcnpg4,secondary}. The results here presented have been obtained with GEANT4, a code originally developed for high-energy physics simulations, but now widely used since many years for a variety of applications, including proton therapy and production of radioisotopes in accelerator-based facilities \cite{g4med,g4medlow}. One of the advantages of GEANT4 is that it allows to choose among different models for the treatment of various physical processes, and different data sets (mostly evaluated libraries) for relevant parameters, such as the reaction cross section. As a consequence, accurate results can be obtained with this code, provided that the most suitable physics list for the studied cases is chosen. 

More details on the Physics List, geometry and material used in the simulations are given in the section \textit{Method}. It is worth mentioning that, for the purpose of this paper, i.e. verifying the importance of a particular reaction in boron-enhanced proton therapy, it is essential that the considered reaction is included in the simulations with a reasonably accurate cross section. This may not always be the case, in particular for low-energy reactions. Physics lists commonly used (including one used by the same authors of Ref. \cite{cirrone:2017} as an example for proton therapy studies \cite{cirrone:2011}) often rely on hadronic models suitable only above hundreds of MeV proton energy. In that case, the cross section at low energy is either missing or highly unreliable. A different approach has therefore to be used. In particular, recent developments have been made in GEANT4 to improve the reliability of the hadronic interactions at very low energy \cite{emilio}.

One of such development regards a high precision model that uses TALYS-based Evaluated Nuclear Data Library (TENDL) for simulating inelastic reactions from 1 to 200 MeV incident energy. More details on TENDL and TALYS can be found in Ref. \cite{tendl1,tendl2} and \cite{talys}, respectively. This option is accessed in GEANT4 through the physic model called \textit{QGSP-BIC-ALLHP}, an extension of the \textit{QGSP-BIC-HP} model, specifically developed to improve the treatment of low-energy nuclear processes. Example of the use of this model can be found in Ref. \cite{g4radio, g4radio2}, where it has been employed to predict isotope production with proton accelerators.

As a first step before performing the simulations, we have verified that the cross section for the p+$^{11}$B$\to$3$\alpha$ reaction used in GEANT4 within the \textit{QGSP-BIC-ALLHP} model is consistent with the experimental one, in the region where they overlap. We have found that this is indeed the case, within a factor of two, as shown in the left panel of Figure \ref{fig:react11B}. However, it should be considered that in TENDL, the cross section starts at 1 MeV proton energy. Clearly, both the lack of data below 1 MeV and a higher cross section above it (relative to experimental data), may somehow affect the reliability of the simulations of the specific reaction under investigation (although by less than a factor of two, in our estimate). However, they do not affect in any way the conclusion of the present work, considering, as will be shown later on, that the proposed mechanism of Proton Boron Capture Therapy would require cross sections (or concentrations) several orders of magnitude higher than presently considered.

Another important aspect that can be investigated with the simulations is the presence of additional reactions that could contribute to explain the decrease in cell survival probability observed in Ref. \cite{cirrone:2017}. In particular, since natural boron was used in the measurement, one may wander whether secondary neutrons produced by proton interaction with oxygen would give rise to the $^{10}$B($n, \alpha$)$^{7}$Li reaction, i.e. the very same one at the basis of Boron Neutron Capture Therapy. In this case, the small neutron-production probability would be compensated by the very high thermal cross section of the reaction, more than three orders of magnitude higher than the p+$^{11}$B one. Neutron production in proton therapy has been extensively studied by means of MC simulations (see for example Ref. \cite{neutrons,neutrons2}), indicating that their contribution to the total dose, together with the one of $\gamma$-rays, is not negligible. It may therefore be reasonable to believe that, in the presence of $^{10}$B in the tissues, they produce an enhancement of the dose to the tissues, with high-LET $\alpha$-particles, as envisaged in BNCT.

\begin{figure}[h!]
 \includegraphics[width=13.cm]{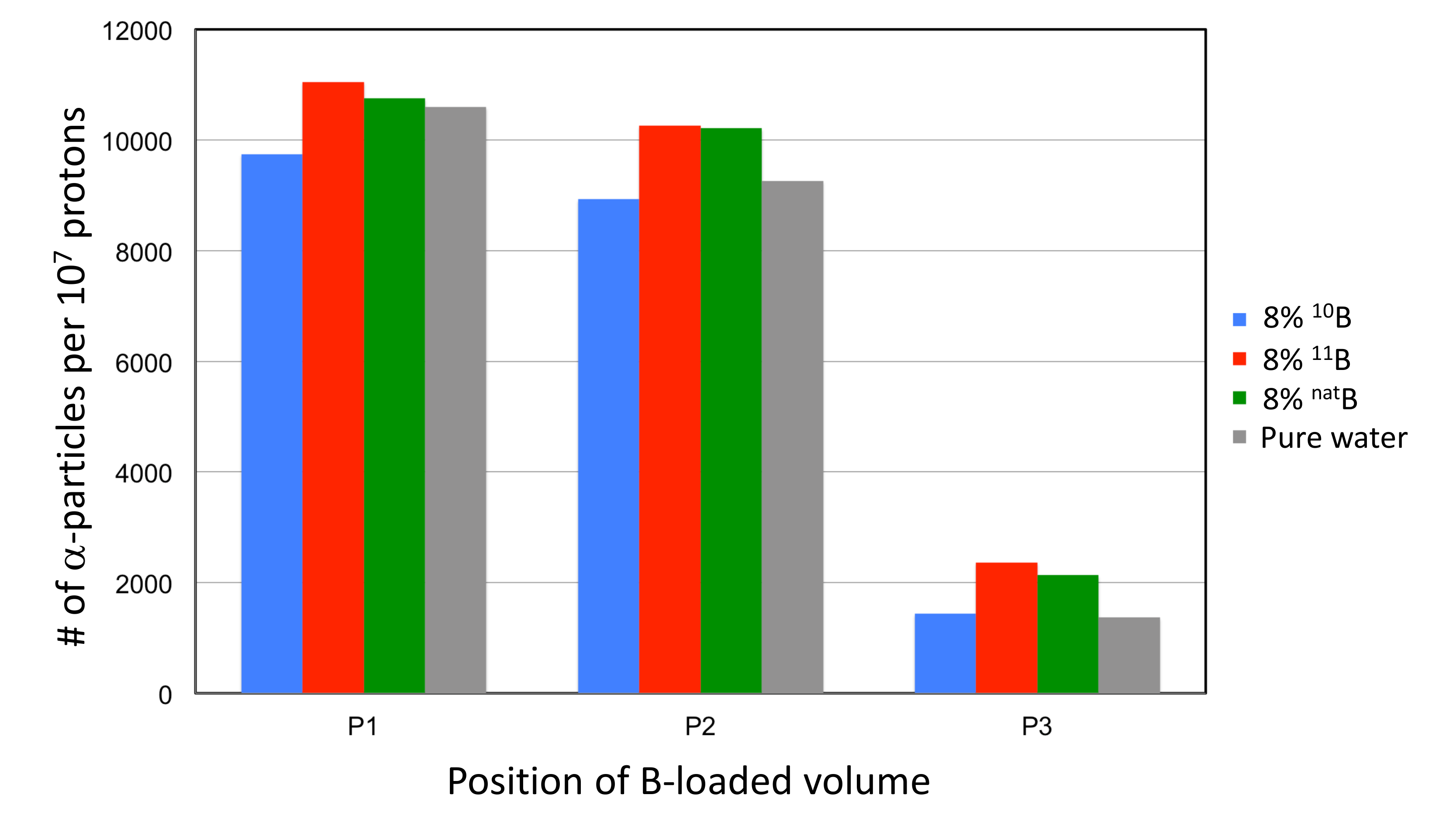}
\caption{Number of produced $\alpha$ particles produced by proton- or neutron-induced reactions on boron, in the three different positions along the SOBP of Ref. \cite{cirrone:2017}. The numbers in the figure are obtained in simulations with and without boron, at concentrations of 8\% (three orders of magnitude higher than realistically achievable), and for 10$^{7}$ incident protons. See text for details. \label{fig:alphas}}
\end{figure} 

Simulations were performed with a proton energy distribution that reproduces the dose distribution, the so-called Spread Out Bragg Peak, of the proton therapy facility of INFN-Laboratori Nazionali del Sud, shown in Ref. \cite{cirrone:2017}, and used in the measurement described therein. 
As a software replica of the cell irradiation experiment described in Ref. \cite{cirrone:2017}, slabs of 1 mm thickness at the entrance of the proton beam (P1 in that reference), and in two positions along the SOBP, at $\sim$24 and $\sim$30 mm (P2 and P3, respectively), were modified so to contain 80 ppm of BSH, with $^{nat}$B. The number of $\alpha$-particles emitted in the slabs was then recorded, and compared with the number obtained without BSH. Furthermore, to discriminate the reaction responsible for their production, additional simulations were performed with 80 ppm of 100\% $^{11}$B and $^{10}$B, respectively. However, it was immediately realized that to obtain a statistically significant result with such a low boron concentration, would have required more than a billion incident protons, implying a huge running time. In fact, while a statistically significant SOBP can be obtained with 10$^{5}$ incident protons, when looking for proton- or neutron-induced reactions on boron, even 10$^{7}$ incident protons do not show any effect. 

A different strategy has therefore to be envisaged in order to introduce a large gain in the interaction probability of proton- and neutron-induced reactions on boron in the simulations. Considering that in GEANT4 it is not possible (or easy) to use biasing techniques, it was decided to perform simulations with a boron concentration increased by a factor of one thousand relative to the realistic case, i.e. to 8\% (instead of 80 ppm that was presumably used in the measurement). The results of these simulations were then scaled down by the same gain factor in the analysis of the dose. Figure \ref{fig:alphas} shows the recorded number of $\alpha$-particles produced in the 1 mm slabs loaded with 8\% of $^{nat}$B, $^{11}$B and $^{10}$B, placed in the three position mentioned above along the proton direction. The results refer to 10$^{7}$ incident protons (more details on the simulations and on the proton incident energy are provided in the \textit{Method} section). For comparison, the number of $\alpha$-particles in pure water are also reported in the figure. Once scaled down to 80 ppm, and normalized for the number of protons, the probability of producing an $\alpha$ particle in a p+$^{nat}$B, or p+$^{11}$B is of the order of 10$^{-7}$. This value is lower than the calculations of the reaction yield presented in the previous subsection, although still consistent with it. It should be considered, in fact, that the calculations represent an upper limit, given the assumptions used for the cross section (the maximum value was adopted for the whole proton energy range in the Bragg peak), and the use of mono-energetic protons, giving rise to a sharp Bragg peak, contrary to the simulations where the Bragg peak is spread out over a large depth. In any case, simulations confirm that the number of proton- or neutron-induced reactions on boron in proton therapy is practically negligible.

An interesting behavior is observed when pure $^{10}$B is loaded in the tissue. In this case, the number of $\alpha$ particles produced in the first two positions even decreases, relative to the pure water case. This behavior could probably be explained by a lower probability of $\alpha$-particle production in direct reactions, relative to $^{11}$B and $^{16}$O. In the position P3, i.e. at the end of the SOBP, a positive contribution from $^{10}$B is also observed, possibly related to reactions induced by thermalized secondary neutrons. 

\begin{figure}[h!]
 \includegraphics[width=13.cm]{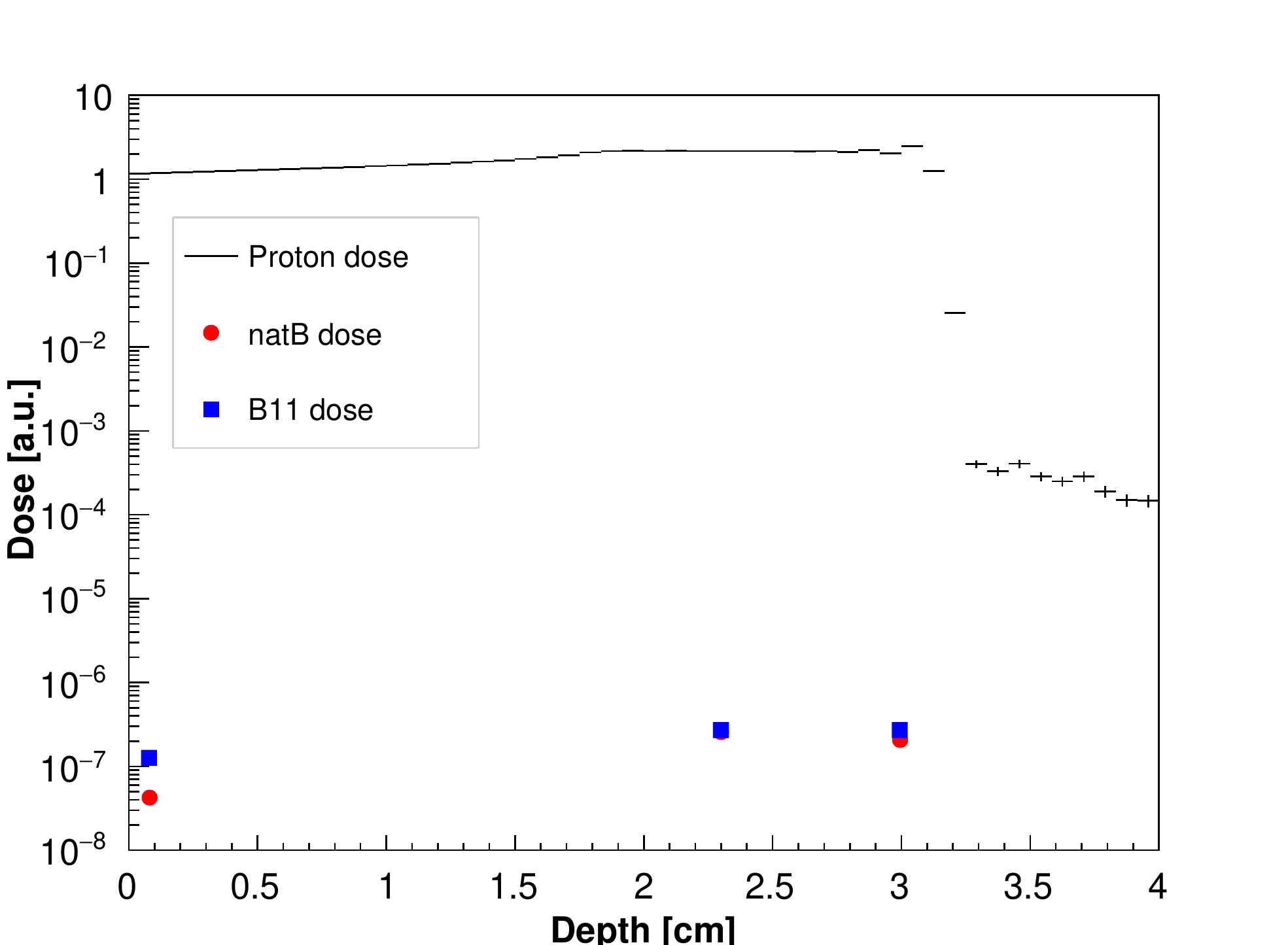}
\caption{The in depth dose profile for a 62 MeV incident proton in a phantom (black curve), compared with the simulated dose related to p+$^{11}$B$\to$$\alpha$ reaction, in 1 mm slabs positioned along the proton direction, containing 80 ppm of natural boron or isotopically pure $^{11}$B. \label{fig:doseg4}}
\end{figure} 

Figure \ref{fig:doseg4} shows the results of the dose profile produced by the proton beam, together with the one produced by $\alpha$ particles in the cases of $^{nat}$B and pure $^{11}$B. These results once more unequivocally indicate that the Proton Boron Capture Therapy does not play a role in the proton therapy, and a different explanation has to be searched for in order to explain the decrease in cell survival probability observed in the measurement of Ref. \cite{cirrone:2017}.

% energy dependence of the cross section, to examine another assumption made by Cirrone \textit{et al.}, i.e. that the p+$^{11}$B reactions occur predominantly in the Bragg peak. The energy-dependence of the cross section was measured by ... in the range from ... to ... MeV. The maximum value of the p+$^{11}$B$\to$3$\alpha$ cross section is 1.2 barn. The same dataset extends to ... MeV, showing a decrease of the cross section to ..00 mbarn. Above that energy no data are available in literature. Nevertheless, it can be reasonably assumed that the cross section does not fall abruptly, but remains of the order  the cross section can be calculated by means of the 
%

\section*{Discussion}

According to calculations and simulations, the expected effect of the p+$^{11}$B$\to$3$\alpha$ reaction is negligible, being the dose related to this reaction orders of magnitude lower than the dose delivered by the primary proton beam. In fact, both results clearly indicate that the reaction probability is too low, at the declared concentration, to bear any effect in the therapy. It may be argued that this comparison holds at the macroscopic level, while at the cellular level the dose ratio may strongly favor the proton-induced reaction on Boron. However, even in this case, it is not clear whether a therapeutic effect would be reached, considering that these reactions would occur both in tumor tissues and in healthy ones, unless a large gain in the boron concentration in tumor is achieved (and at concentrations orders of magnitude higher than reasonably achievable). We remind here that, as in the case of BNCT, the use of BSH or similar boron carries (such as BoronoPhenylAlanine, BPA) can only lead to a small gain of boron uptake in the tumor tissues, relative to the healthy ones. Therefore it seems highly unlikely that the decrease in cellular survival probability and larger cellular damage reported by Cirrone \textit{et al.} in Ref. \cite{cirrone:2017} in the presence of BSH is related with the p+$^{11}$B$\to$3$\alpha$, or to any other proton- or neutron-induced reaction on boron (being neutrons produced as secondary particles by the proton interaction in tissues). Different explanations have therefore to be looked for, such as, for example, an unreported much higher boron concentration in the cell specimen. Another possibility would be the effect of BSH in increasing the radio-sensitivity of the cells, a purely biochemical effect, that has little to do with Nuclear Physics. More studies would be required to establish the exact origin of the observed higher effectiveness of the proton therapy in the presence of BSH, if any, and concluding on the final applicability of the technique. 

\section*{Methods}

In this section, some details on the analytical calculations and Monte Carlo simulations performed with the GEANT4 toolkit are reported. 

\subsection*{Calculation of the boron areal density}

From the BSH stoichiometric formula the number of atoms is: 2$\times$Na (weight 23), 12$\times$B (weight 11), 12$\times$H (weight 1), and S (weight 32). The atomic weights were  approximated to integer numbers of atomic mass units, being the tiny difference ininfluent on the overall result. Known the Avogadro's number, one can immediately find the number of water molecules (weight 18) in 1 ml as 3.35$\times$10$^{22}$; from the 0.17 mg/ml BSH concentration one finds the number of BSH molecules as 4.61$\times$10$^{17}$; with 12 boron atoms per BSH molecules, the number of boron atoms per water molecule is 1.65$\times$10$^{-4}$. 
The areal density for a 1 mm thickness of such a sample is thus given by the number of boron atoms per molecule (12) times the number of BSH molecules per cm$^{3}$ (4.61$\times$10$^{17}$) times the sample thickness (0.1 cm): 
n$_{B}$ = 12$\times$4.61$\times$10$^{17}\times$0.1 = 5.53$\times$10$^{17}$ cm$^{-2}$ or equivalently 5.5$\times$10$^{-7}$ atoms/barn.

\subsection*{Geometry and Physics List used in GEANT4 simulations}

In GEANT4, a software replica of the phantom was implemented, consisting of a volume of 2$\times$2$\times$10 cm$^{3}$ of water. The specific characteristics of the simulation required, for generating the incident primary particles, the use of the G4GeneralParticleSource package, which allows one to specify the spatial, angular and energy distribution of the incident particles. In the present work, a pencil beam was used, hitting the phantom perpendicular to the input face. For the spatial distribution, a Gaussian beam of 2 mm sigma was used. The beam travels along the axis of the phantom volume without any angular dispersion. The energy distribution of the proton beam was suitably constructed in order to obtain a SOBP as closer as possible to the SOBP reported in Ref. \cite{cirrone:2017}. To this purpose, a user-defined histogram of a given number of bins in a given energy range is defined, by providing the upper boundary of each bin and a weight. In this work, the range was chosen between 45 and 60 MeV proton energy. The weights in the various bins were varied so to obtain the experimental SOBP, in the same way as reported in an example suggested by the University of New Mexico Particle Therapy Group \cite{unm}. 

In all simulations presented in this work, 10$^{7}$ primary protons are generated. A particular care is devoted to the choice of the physic processes to be simulated. The adopted physic list is the same one used for the hadrontherapy example in Ref. \cite{unm}, 
except that the G4HadronPhysicsQGSP\_BIC\_AllHP module replaced the G4HadronPhysicsQGSP\_BIC\_HP one. Inelastic nuclear interactions were described by a binary cascade model followed by a pre-compound model of nuclear de-excitation. It is similar to the QGSP\_BIC model, except that for neutrons of 20 MeV and lower the High Precision neutron models and cross sections are used to describe elastic and inelastic scattering, capture and fission. For protons, deuterons and alpha particles below 100 MeV the High Precision cross sections for elastic and inelastic scattering are used, which are based on the TENDL nuclear data library (TALYS-based Evaluated Nuclear Data Library) for isotope production \cite{emilio}. 
Electromagnetic interactions were described by the standard model (G4EmStandardPhysics\_option4), while elastic nuclear interactions are described by the high precision hadronic elastic physics model (G4HadronElasticPhysicsHP). Additionally, theG4EmExtraPhysics, G4StoppingPhysics, G4IonBinaryCascadePhysics physics modules were used. The range threshold for secondary particle production was set to 0.1 mm.

In three different positions in the phantom, along the incident proton beam direction, a slab of 1 mm thickness of water is enriched with either natural boron, $^{11}$B and $^{10}$B isotopes. The slabs are positioned at distances from the phantom surface of 0.2 cm (P1), 2.4 cm (P2) and 3 cm (P3). 

\bibliography{pbct_arx}

\begin{thebibliography}{10}
\expandafter\ifx\csname url\endcsname\relax
  \def\url#1{\texttt{#1}}\fi
\expandafter\ifx\csname urlprefix\endcsname\relax\def\urlprefix{URL }\fi
\expandafter\ifx\csname doiprefix\endcsname\relax\def\doiprefix{DOI }\fi
\providecommand{\bibinfo}[2]{#2}
\providecommand{\eprint}[2][]{\url{#2}}

\bibitem{yoon:2014}
\bibinfo{author}{Yoon, D.-K.}, \bibinfo{author}{Jung, J.-Y.} \&
  \bibinfo{author}{Suh, T.~S.}
\newblock \bibinfo{journal}{\bibinfo{title}{Application of proton boron fusion
  reaction to radiation therapy: A monte carlo simulation study.}}
\newblock {\emph{\JournalTitle{APPLIED PHYSICS LETTERS}}}
  \textbf{\bibinfo{volume}{105}}, \bibinfo{pages}{223507}
  (\bibinfo{year}{2014}).

\bibitem{mcnp}
\bibinfo{author}{Waters, L.}, \bibinfo{author}{Hendricks, J.} \&
  \bibinfo{author}{McKinney, G.}
\newblock \bibinfo{journal}{\bibinfo{title}{Mcnpx 2.6.0. users's guide}}.
\newblock {\emph{\JournalTitle{Los Alamos National Laboratory}}}
  (\bibinfo{year}{2008}).

\bibitem{yoon:2017}
\bibinfo{author}{Jung, J.-Y.} \emph{et~al.}
\newblock \bibinfo{journal}{\bibinfo{title}{Comparison between proton boron
  fusion therapy (pbft) and boron neutron capture therapy (bnct): a monte carlo
  study.}}
\newblock {\emph{\JournalTitle{Oncotarget}}} \textbf{\bibinfo{volume}{8}},
  \bibinfo{pages}{39774--39781} (\bibinfo{year}{2017}).

\bibitem{cirrone:2017}
\bibinfo{author}{Cirrone, G. A.~P.} \emph{et~al.}
\newblock \bibinfo{journal}{\bibinfo{title}{First experimental proof of proton
  boron capture therapy (pbct) to enhance protontherapy effectiveness modelling
  pet radionuclide production in tissue and external targets using geant4.}}
\newblock {\emph{\JournalTitle{Scientific Reports}}}
  \textbf{\bibinfo{volume}{8}}, \bibinfo{pages}{1141} (\bibinfo{year}{2018}).

\bibitem{bnct}
\bibinfo{author}{Nedunchezhian, K.}, \bibinfo{author}{Aswath, N.},
  \bibinfo{author}{Thiruppathy, M.} \& \bibinfo{author}{Thirugnanamurthy, S.}
\newblock \bibinfo{journal}{\bibinfo{title}{Boron neutron capture therapy - a
  literature review}}.
\newblock {\emph{\JournalTitle{J. Clin. Diagn. Res.}}}
  \textbf{\bibinfo{volume}{10}}, \bibinfo{pages}{ZE01–ZE04}
  (\bibinfo{year}{2016}).

\bibitem{geant4}
\bibinfo{author}{Agostinelli, S.} \emph{et~al.}
\newblock \bibinfo{journal}{\bibinfo{title}{Geant4, a simulation toolkit.}}
\newblock {\emph{\JournalTitle{Nucl. Instr. Meth. A}}}
  \textbf{\bibinfo{volume}{506}}, \bibinfo{pages}{250--303}
  (\bibinfo{year}{2003}).

\bibitem{exfor}
\urlprefix\url{https://www-nds.iaea.org/exfor/exfor.htm}.

\bibitem{segel}
\bibinfo{author}{Segel, R.}, \bibinfo{author}{Hanna, S.} \&
  \bibinfo{author}{Allas, R.}
\newblock \bibinfo{journal}{\bibinfo{title}{States in c12 between 16.4 and 19.6
  mev}}.
\newblock {\emph{\JournalTitle{Physical Review B}}}
  \textbf{\bibinfo{volume}{139}}, \bibinfo{pages}{818} (\bibinfo{year}{1965}).

\bibitem{neutrons}
\bibinfo{author}{Jia, S.~B.}, \bibinfo{author}{Hadizadeha, M.~H.},
  \bibinfo{author}{Mowlavi, A.~A.} \& \bibinfo{author}{Loushaba, M.~E.}
\newblock \bibinfo{journal}{\bibinfo{title}{Evaluation of energy deposition and
  secondary particle production in proton therapy of brain using a slab head
  phantom.}}
\newblock {\emph{\JournalTitle{Reports of Practical Oncology and
  Radiotherapy}}} \textbf{\bibinfo{volume}{19}}, \bibinfo{pages}{376–384}
  (\bibinfo{year}{2014}).

\bibitem{neutrons2}
\bibinfo{author}{Dawidowska, A.}, \bibinfo{author}{Paluch~Ferszt, M.} \&
  \bibinfo{author}{Konefał, A.}
\newblock \bibinfo{journal}{\bibinfo{title}{The determination of a dose
  deposited in reference medium due to (p,n) reaction occurring during proton
  therapy.}}
\newblock {\emph{\JournalTitle{Reports of Practical Oncology and
  Radiotherapy}}} \textbf{\bibinfo{volume}{19}}, \bibinfo{pages}{53--58}
  (\bibinfo{year}{2014}).

\bibitem{oxy}
\bibinfo{author}{Gruhle, W.} \& \bibinfo{author}{Kober, B.}
\newblock \bibinfo{journal}{\bibinfo{title}{The reactions $^{16}$o($p,
  \alpha$), $^{20}$ne($p, \alpha$), and $^{24}$mg($p, \alpha$)}}.
\newblock {\emph{\JournalTitle{Nucl. Phys. A}}} \textbf{\bibinfo{volume}{286}},
  \bibinfo{pages}{523} (\bibinfo{year}{1977}).

\bibitem{tendl1}
\bibinfo{author}{Koning, A.} \emph{et~al.}
\newblock \bibinfo{title}{Tendl-2015: Talys-based evaluated nuclear data
  library}.
\newblock \urlprefix\url{https://tendl.web.psi.ch/tendl_2015/tendl2015.html}.

\bibitem{tendl2}
\bibinfo{author}{Koning, A.} \& \bibinfo{author}{Rochman, D.}
\newblock \bibinfo{journal}{\bibinfo{title}{Modern nuclear data evaluation with
  the talys code system}}.
\newblock {\emph{\JournalTitle{Nuclear Data Sheets}}}
  \textbf{\bibinfo{volume}{113}}, \bibinfo{pages}{2841} (\bibinfo{year}{2012}).

\bibitem{talys}
\bibinfo{author}{Koning, A.}, \bibinfo{author}{Hilaire, S.} \&
  \bibinfo{author}{Duijvestijn, M.}
\newblock \bibinfo{journal}{\bibinfo{title}{Talys-1.0}}.
\newblock {\emph{\JournalTitle{Proceedings of the International Conference on
  Nuclear Data for Science and Technology - ND2007, April 22-27, 2007, Nice,
  France, eds. O. Bersillon, F. Gunsing, E. Bauge, R. Jacqmin and S. Leray, EDP
  Sciences}}} \bibinfo{pages}{211--214} (\bibinfo{year}{2008}).

\bibitem{fluka}
\bibinfo{author}{Battistoni, G.} \emph{et~al.}
\newblock \bibinfo{journal}{\bibinfo{title}{The fluka code and its use in
  hadron therapy}}.
\newblock {\emph{\JournalTitle{Il Nuovo Cimento C}}}
  \textbf{\bibinfo{volume}{31}}, \bibinfo{pages}{69--75}
  (\bibinfo{year}{2008}).

\bibitem{mcnpg4}
\bibinfo{author}{Titt, U.}, \bibinfo{author}{Bednarz, B.} \&
  \bibinfo{author}{Paganetti, H.}
\newblock \bibinfo{journal}{\bibinfo{title}{Comparison of mcnpx and geant4
  proton energy deposition predictions for clinical use}}.
\newblock {\emph{\JournalTitle{Phys. Med. Biol.}}}
  \textbf{\bibinfo{volume}{57}}, \bibinfo{pages}{6381--6393}
  (\bibinfo{year}{2012}).

\bibitem{secondary}
\bibinfo{author}{Robert, C.} \emph{et~al.}
\newblock \bibinfo{journal}{\bibinfo{title}{Distributions of secondary
  particles in proton and carbon-ion therapy: a comparison between gate/geant4
  and fluka monte carlo codes.}}
\newblock {\emph{\JournalTitle{Phys. Med. Biol.}}}
  \textbf{\bibinfo{volume}{58}}, \bibinfo{pages}{2879--2899}
  (\bibinfo{year}{2013}).

\bibitem{g4med}
\bibinfo{author}{McKinnon, S.} \emph{et~al.}
\newblock \bibinfo{journal}{\bibinfo{title}{Local dose enhancement of proton
  therapy by ceramic oxide nanoparticles investigated with geant4
  simulations.}}
\newblock {\emph{\JournalTitle{Physica Medica}}} \textbf{\bibinfo{volume}{32}},
  \bibinfo{pages}{1584--1593} (\bibinfo{year}{2016}).

\bibitem{g4medlow}
\bibinfo{author}{Ratcliffe, N.}, \bibinfo{author}{Barlow, R.},
  \bibinfo{author}{Bungau, A.}, \bibinfo{author}{Bungau, C.} \&
  \bibinfo{author}{Cywinski, R.}
\newblock \bibinfo{journal}{\bibinfo{title}{Geant4 target simulations for low
  energy medical applications.}}
\newblock {\emph{\JournalTitle{Proceedings of the 4th International Particle
  Accelerator Conference IPAC2013, Shangai, China}}}
  \textbf{\bibinfo{volume}{130}}, \bibinfo{pages}{3717--3719}
  (\bibinfo{year}{2013}).

\bibitem{cirrone:2011}
\bibinfo{author}{Cirrone, G.} \emph{et~al.}
\newblock \bibinfo{journal}{\bibinfo{title}{Hadrontherapy: a geant4-based tool
  for proton/ion-therapy studies}}.
\newblock {\emph{\JournalTitle{Progress in Nuclear Science and Technology}}}
  \textbf{\bibinfo{volume}{2}}, \bibinfo{pages}{207--211}
  (\bibinfo{year}{2011}).

\bibitem{emilio}
\bibinfo{author}{Mendoza, E.}, \bibinfo{author}{Sansaloni, F.},
  \bibinfo{author}{Arce, P.}, \bibinfo{author}{Cano-Ott, D.} \&
  \bibinfo{author}{Lagares, J.~I.}
\newblock \bibinfo{journal}{\bibinfo{title}{A new physics model for the charged
  particle transport with geant4}}.
\newblock {\emph{\JournalTitle{2011 IEEE Nuclear Science Symposium Conference
  Record}}} \bibinfo{pages}{2242--2244} (\bibinfo{year}{2011}).

\bibitem{g4radio}
\bibinfo{author}{Poignant, F.}, \bibinfo{author}{Penfold, S.},
  \bibinfo{author}{Asp, J.}, \bibinfo{author}{Takhar, P.} \&
  \bibinfo{author}{Jackson, P.}
\newblock \bibinfo{journal}{\bibinfo{title}{Geant4 simulation of cyclotron
  radioisotope production in a solid target}}.
\newblock {\emph{\JournalTitle{Physica Medica}}} \textbf{\bibinfo{volume}{32}},
  \bibinfo{pages}{728--734} (\bibinfo{year}{2016}).

\bibitem{g4radio2}
\bibinfo{author}{Amin, T.}, \bibinfo{author}{Infantino, A.},
  \bibinfo{author}{Lindsay, C.}, \bibinfo{author}{Barlow, R.} \&
  \bibinfo{author}{Hoehr, C.}
\newblock \bibinfo{journal}{\bibinfo{title}{Modelling pet radionuclide
  production in tissue and external targets using geant4}}.
\newblock {\emph{\JournalTitle{Physica Medica}}} \textbf{\bibinfo{volume}{32}},
  \bibinfo{pages}{728--734} (\bibinfo{year}{2016}).

\bibitem{unm}
\bibinfo{author}{Keyes, R.~W.}, \bibinfo{author}{Romano, C.},
  \bibinfo{author}{Arnold, D.} \& \bibinfo{author}{Luan, S.}
\newblock \bibinfo{journal}{\bibinfo{title}{Radiation therapy calculations
  using an on-demand virtual cluster via cloud computing}}.
\newblock {\emph{\JournalTitle{arXiv:1009.5282}}}  (\bibinfo{year}{2010}).
\newblock \urlprefix\url{http://arxiv.org/abs/1009.5282}.

\end{thebibliography}

\end{document}